\documentclass[pra,twocolumn,showpacs]{revtex4}
\usepackage{epsfig}
\usepackage{amsbsy}
\usepackage{amsmath}
\usepackage{latexsym}

\begin{document}

\title{Thermal Entanglement in Ferrimagnetic Chains}
\author{Xiaoguang Wang$^{1,2}$ and Z. D. Wang$^{1,3}$}
\affiliation{ $^{1}$ Department of Physics and Center of
Theoretical and Computational Physics, The University of Hong
Kong, Pokfulam Road, Hong Kong, China.\\
 $^{2}$Zhejiang Institute
of Modern Physics, Department of Physics, Zhejiang University,
Hanzhou 310027, China.\\
$^{3}$ National Laboratory of Solid State Microstructures, Nanjing
University, Nanjing, China }
\date{\today}
\begin{abstract}
A formula to evaluate the entanglement in an one-dimensional
ferrimagnetic system is derived. Based on the formula, we find that
the thermal entanglement in a small size spin-1/2 and spin-$s$
ferrimagnetic chain  is rather robust against temperature, and the
threshold temperature may be arbitrarily high when $s$ is
sufficiently large.
This intriguing result answers
unambiguously
 a fundamental question:
``can entanglement and quantum behavior in physical systems
survive at arbitrary high temperatures?"

\end{abstract}
\pacs{03.65.Ud, 75.10.Jm }
\maketitle

A physical system may exhibit entanglement at a finite
temperature~\cite{M_Nielsen,M_Arnesen,M_Wang01,M_Gunlycke,Vedral}.
 The thermal
entanglement always vanishes above a threshold temperature for
systems with finite Hilbert space dimension~\cite{Fine}. Recently,
Ferreira {\it et al.} raised a fundamental question: can
entanglement and quantum behavior in physical systems survive at
arbitrary high temperatures?~\cite{Ferreira} They found that the
entanglement between a cavity mode and a movable mirror does occur
for any finite temperature. This result sheds a new light on the
question and help to understand macroscopic properties of solids.

In this report, we derive a formula to evaluate the entanglement
in an one-dimensional ferrimagnetic system. Intriguingly, we find
that the entanglement is rather robust against temperature in this
kind of system, with the Hamiltonian
\begin{equation}
H=J\sum_n ({\bf s}_n\cdot{\bf S}_{n+1}+{\bf S}_{n+1}\cdot {\bf
s}_{n+2}),\label{H1}
\end{equation}
where ${\bf s}_n$ and ${\bf S}_n$ are spin-1/2 and spin-$s$
operators, respectively. The antiferromagnetic exchange
interactions exist only between nearest neighbors, and they are of
the same strength which is set to unity $(J=1)$. Physically, the
system contains two kinds of spins, spin $\frac{1}{2}$ and $s$,
alternating on a ring (or a chain with the periodic boundary
condition).

Let us now study the entanglement of states of the system at
thermal equilibrium described by the density operator
$\rho(T)=\exp(-\beta H)/Z$, where $\beta=1/k_BT$, $k_B$ is the
Boltzmann's constant, which is assume to be 1, and
$Z=\text{Tr}\{\exp(-\beta H)\}$ is the partition function. The
entanglement in the thermal state is referred to as the thermal
entanglement.

To study quantum entanglement in the ferrimagnetic system, we need
a good entanglement measure. One possible way is to use the
negativity~\cite{Vidal} based on the partial transpose
method~\cite{PH}. In the cases of two spin halves and the (1/2,1)
mixed spins, a positive partial transpose (PPT) (or the non-zero
negativity) is necessary and sufficient for separability
(entanglement). Although the present ferrimagnetic system is a
kind of (1/2,$s$) system, fortunately, it was shown that due to
the SU(2) symmetry in the model Hamiltonian (\ref{H1}), the
non-zero negativity is still a necessary and sufficient condition
for entanglement between a spin half and spin
$s$~\cite{Schliemann}. This result allows us to {\it exactly}
investigate entanglement features of our mixed spin systems.

The negativity of a state $\rho$ is defined as
\begin{equation}
{\cal N(\rho)}=\sum_i|\mu_i|,
\end{equation}
where $\mu_i$ is the negative eigenvalue of $\rho^{T_2}$, and $T_2$ denotes \\
the partial transpose with respect to the second system. The
negativity ${\cal N}$ is related to the trace norm of $\rho^{T_2}$
via
\begin{equation}
{\cal N(\rho)}=\frac{\|\rho^{T_2}\|_1-1}{2},
\end{equation}
where the trace norm of $\rho^{T_2}$ is equal to the sum of the
absolute values of the eigenvalues of $\rho^{T_2}$.

Obviously, our system has the SU(2) symmetry, and any two-spin
reduced density matrix from the thermal state is also
SU(2)-invariant. Now, we consider the entanglement between the
spin half and spin $s$, and derive the corresponding expression of
negativity by the partial time reversal method~\cite{breuer},
which is equivalent to the partial transpose method up to a local
unitary operator.

The density matrix of an SU(2)-invariant state for the spin half
and spin $s$ can be written in the form
\begin{equation}
\rho=\frac{F}{2s}{\bf P}_{s-1/2}+\frac{1-F}{2s+2}{\bf P}_{s+1/2}
\end{equation}
with $F\in [0,1]$. One may immediately check that the parameter
$F$ is identical to the expectation value of the projector ${\bf
P}_{s-1/2}$ on the densitry matrix, i.e., $F=\langle {\bf
P}_{s-1/2} \rangle$.
Noting the fact that ${\bf
P}_{s-1/2}+ {\bf P}_{s+1/2}=1$, we rewrite the density operator as
\begin{equation}\label{su2}
\rho=\frac{(2sF+F-s){\bf P}_{s-1/2}}{2s(s+1)}+\frac{1-F}{2s+2}.
\end{equation}
and obtain the relations
\begin{align}\label{ppp}
{\bf P}_{s-1/2}=&\frac{1}{2s+1}(s-2\mathbf{s}_1\cdot
\mathbf{S}_2)\notag\\
{\bf P}_{s+1/2}=&\frac{1}{2s+1}(s+1+2\mathbf{s}_1\cdot
\mathbf{S}_2).
\end{align}
The partial time reversal operator $\tau_2$ changes the sign of
${\bf S}_2$ and ${\bf s}_1\cdot{\bf S}_2$
\begin{align}
\tau_2 ({\bf S}_2)=&-{\bf S}_2\notag\\
\tau_2 ({\bf s}_1\cdot{\bf S}_2)=&-{\bf s}_1\cdot{\bf S}_2,
\end{align}
Therefore, we get
\begin{equation}\label{taop}
\tau_2 ({\bf P}_{s-1/2})=\frac{2s}{2s+1}-{\bf P}_{s-1/2}.
\end{equation}
We see that the partial time reversal also changes the sign of
projector, but with an additional additive constant.

Using Eqs.~(\ref{su2}) and (\ref{taop}), the partial time reversed
density matrix is obtained as
\begin{equation}
\rho^{\tau_2}=-\frac{(2sF+F-s){\bf
P}_{s-1/2}}{2s(s+1)}+\frac{2sF+F+1}{2(2s+1)(s+1)}.
\end{equation}
Any projector has two eigenvalues 0 and 1. Thus, from the above
equation, we deduce that only the following eigenvalue of
$\rho^{\tau_2}$ is possibly negative
\begin{align}
\lambda=&\frac{1}{2s+1}-\frac{F}{2s}\notag\\
=&\frac{1}{2s+1}\left(\frac{1}2+\frac{1}s \langle{\bf
s}_1\cdot{\bf S}_2\rangle\right),
\end{align}
where we have used the first equality in Eq.~(\ref{ppp}). By
taking into account that the eigenvalue occurs with multiplicity
$2s$, we finally obtain the negativity
\begin{equation}\label{NNN}
\mathcal{N}=\max \left[ 0,-\frac{s+2\langle \mathbf{s}_1\cdot \mathbf{S}%
_2\rangle }{2s+1}\right] .
\end{equation}
The negativity is only determined by a single correlator, which is
due to the fact that our state is SU(2)-invariant.

At this stage, to obtain straightforwardly analytical results of
negativity in the present ferrimagnetic system as well as to gain
some essential physical insight into entanglement features
like the intriguing high-temperature entanglement, we here focus
only on a two-spin case first. In this case, the Heisenberg
interaction can be written in terms of projectors as follows
\begin{align}\label{expand}
H=\mathbf{s}_1\cdot \mathbf{S}_2=&\frac 12[(\mathbf{s}_1+\mathbf{S}_2)^2-%
\mathbf{s}_1^2-\mathbf{S}_2^2]  \nonumber \\
=&\frac 12(\mathbf{s}_1+\mathbf{S}_2)^2-\frac
38-\frac{s(s+1)}2\notag\\
=&-\frac{s+1}{2}{\bf P}_{s-1/2}+\frac{s}2 {\bf P}_{s+1/2}.
\end{align}

From Eq.~(\ref{expand}), the eigenvalues are simply
\begin{equation}\label{energy}
E_0=-\frac{s+1}2, E_1=\frac s2.
\end{equation}
The eigenvalue is just the expectation value of the correlator $
\langle {\bf s}_1\cdot {\bf S}_2\rangle$ in the corresponding
state. From Eqs.~(\ref{energy}) and (\ref{NNN}), the negativity
for the ground state and the first excited state are obtained as
\begin{equation}
\mathcal{N}_0=\frac 1{2s+1} ,\; \mathcal{N}_1=0.
\end{equation}
Clearly, the ground-state is entangled, while the first-excited
state is not. In addition, we observe that the ground-state
entanglement decreases as $s$ increases, and finally vanishes in
the limit of $s=\infty$. Note that, the case of $s=\infty$
corresponds actually to a classical spin, and thus, as expected
intuitively, there exists no entanglement between a quantum system
and a classical one indeed.

\begin{figure}
\includegraphics[width=0.45\textwidth]{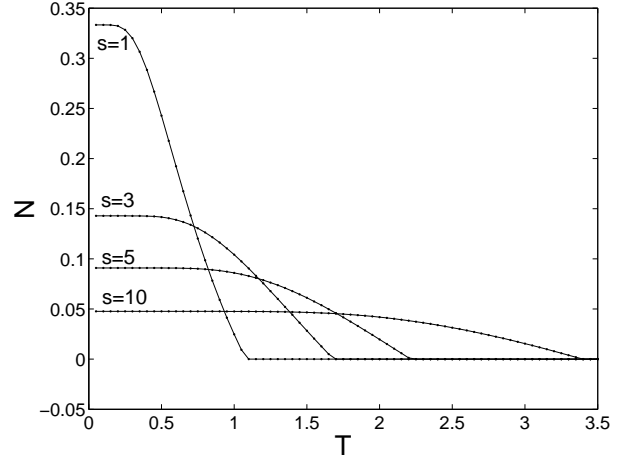}
\caption{Negativity versus temperature for different $s$ in the
two-site system.}
\end{figure}

At finite temperatures, we need to know the correlator $\langle
\mathbf{s}_1\cdot \mathbf{S}_2\rangle$ evaluated on the thermal
state, which completely determines the negativity. From
Eq.~(\ref{energy}),  we obtain readily the partition function and
the correlator as
\begin{equation}\label{pfun}
Z=2se^{\frac{(s+1)\beta}2}+2(s+1)e^{-\frac {s\beta}2}
\end{equation}
and 
\begin{equation}\label{correlator}
\langle \mathbf{s}_1\cdot \mathbf{S}_2\rangle =-\frac 1Z\frac{\partial Z}{%
\partial \beta }=-\frac{s(s+1)}Z\left( e^{\frac{{s+1}\beta}2}-e^{-\frac
{s\beta}2}\right) .
\end{equation}
Substituting Eqs.~(\ref{correlator}) and (\ref{pfun}) into
Eq.~(\ref{NNN}),  the negativity is explicitly expressed as
\begin{equation}\label{NNNN}
{\cal
N}=\frac{2s}{(2s+1)Z}\max\{0,e^{\frac{(s+1)\beta}2}-2(s+1)e^{-\frac{s\beta}2}\}
\end{equation}
The numerical results of negativity versus temperature for
different $s$ are plotted in Fig.~1. As expected, for a fixed $s$,
the negativity decreases monotonically as  temperature increases,
and vanishes when the temperature is equal or larger than a
threshold value $T_\text{th}$. But more arrestingly,  when $s$ is
large, the entanglement is rather robust against temperature.

\begin{figure}
\includegraphics[width=0.45\textwidth]{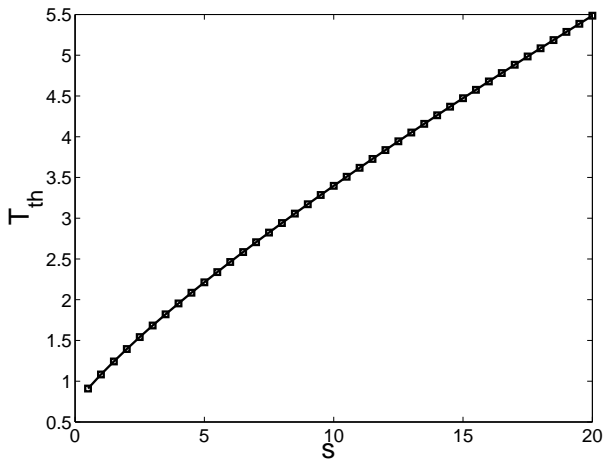}
\caption{Threshold temperature versus $s$.}
\end{figure}

From Eq.~(\ref{NNNN}), the threshold value of the thermal
entanglement is found to be
\begin{equation}
T_{\text{th}}=\frac{2s+1}{2\ln (2s+2)},
\end{equation}
from which we immediately have a striking result
\begin{equation}
\lim_{s\rightarrow\infty} T_\text{th}=\infty.
\end{equation}
The threshold temperature $T_\text{th}$ can be arbitrarily high
when $s$ is large enough. A plot of $T_\text{th}$ versus $s$ is
illustrated in Fig.~2, and we observe that the threshold
temperature increases with the increase of $s$.

\begin{figure}
\includegraphics[width=0.45\textwidth]{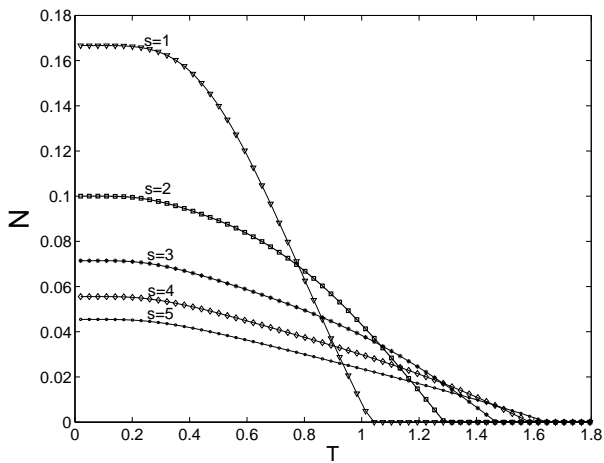}
\caption{Negativity versus temperature for different $s$ in the
four-site system.}
\end{figure}

 It is now interesting to estimate some typical values in terms of
 Eq. (18) (with the unit $J/k_B$).
The value of exchange constant $J$ is of the order $meV$. For a
temperature $T\sim 127 K$, the corresponding $s$ is estimated as
$s=50$, above which the thermal entanglement exists. In this case,
the negativity is approximately 0.01 at temperatures well below
the above temperature.

 Finally, we wish to address briefly the negativity of the systems
 with a larger number of sites $L$.
   For the
 four-site case, we can still have exact results and plot the
 negativity versus temperature for different $s$ in Fig. 3. It is
 seen that
 the negativity behaviors of the four-site case are qualitatively the same as
 those of the two-site case, namely, when $s$ increases, the negativity at zero
 temperature decreases and the threshold temperature increases.
 As for  even larger sizes $L=2N$, due to the limitation of computational resource,
 we here can only make a qualitative analysis with the help of  an approximate analytical
 result  for the ground state energy obtained from
 the spin-wave theory~\cite{Pati}.
 At zero temperature, the nearest spin correlator~\cite{Pati}
 $\langle
\mathbf{s}_1\cdot \mathbf{S}_2\rangle_0=E_0(L)/2L=-s/2-\delta
(s)$, where the subscript index $0$ denotes the ground state, and
$\delta (s)>0$
 and approaches (1/4) in the limit $s \rightarrow  \infty$.
Therefore, from Eq.(11), the negativity
 $\mathcal{N}=2\delta (s)/(2s+1)$ is non-zero at zero temperature.
 Also, for a small size chain,  we note that there exists
 a energy gap (proportional to $s$)
 between the excited
 and ground states, so the negativity is mainly determined
by the ground state contribution at finite temperatures well below
the gap energy, and may be non-zero at high temperature for a very
large $s$.

In conclusion, by considering a simple ferrimagnetic chain model,
we have found that the entanglement is rather robust against
temperature. As the spin $s$ increases, the threshold temperature
for entanglement can be arbitrarily high for a small size chain.
We hope that the present work motivates interests to
investigate other physical systems which display high-temperature
entanglement.

\acknowledgments The authors thank  Y. Chen, F. C. Zhang, and G. M.
Zhang for helpful discussions. X. Wang was supported by NSF-China
under grant no. 10405019, Specialized Research Fund for the Doctoral
Program of Higher Education (SRFDP) under grant No.20050335087, and
the project-sponsored by SRF for ROCS and SEM. Z. D. Wang was
supported by the RGC grant of Hong Kong under No. HKU7045/05P, the
URC fund of HKU, and NSF-China under grant no.  10429401.


\begin{thebibliography}{99}

\bibitem{M_Nielsen}  M. A. Nielsen, Ph. D thesis, University of
Mexico, 1998, quant-ph/0011036.
\bibitem{M_Arnesen}M. C. Arnesen, S. Bose, and V. Vedral, Phys. Rev. Lett. {\bf 87}, 017901 (2001).
\bibitem{M_Wang01}
X. Wang, Phys. Rev. A {\bf 64}, 012313 (2001); Phys.  Lett. A,{\bf
281},  101 (2001).
\bibitem{M_Gunlycke}
D. Gunlycke, V. M. Kendon, V. Vedral, and S. Bose, \pra {\bf 64},
042302 (2001).
\bibitem{Vedral}V. Vedral, New J. Phys. {\bf 6}, 102 (2004).
\bibitem{Fine}V. Fine, F. Mintert, and A. Buchleitner, Phys. Rev.
B {\bf 71}, 153105 (2005).
\bibitem{Ferreira}A. Ferreira, A. Guerreiro, and V. Vedral,
quant-ph/0504186.

\bibitem{Vidal}G. Vidal and R. F. Werner, \pra {\bf 65}, 032314
(2002).

\bibitem{PH}A. Peres, \prl {\bf 77}, 1413 (1996);
M. Horodecki, P. Horodecki, and R. Horodecki, Phys. Lett. A {\bf
223}, 1 (1996).

\bibitem{Schliemann}J. Schliemann, \pra {\bf 68}, 012309 (2003).

\bibitem{breuer} H. P. Breuer, Phys. Rev. A \textbf{71}, 062330
(2005).

\bibitem{Pati} S. K. Pati, S. Ramaseha, and D. Sen, Phys. Rev. B \textbf{55}, 8894 (1997).

\end{thebibliography}
\end{document}